\documentclass[12pt,a4paper]{article} 
\begin{document}
\begin{center}
{\bf \Large Characterising 
Extrasolar Planets in Reflected Light and Thermal Emission \footnote{published
in {\it SF2A: Scientific Highlights 2002}. Eds. F. Combes and D. Barret. EDPScience, 2002, p. 563}} \\

\bigskip

{\large \it Schneider, J.\\
CNRS - Observatoire de Paris, 92195 Meudon, France - Jean.Schneider@obspm.fr}
\end{center}

\mbox{}\vspace{1.cm}\\

{\bf Abstract}
 
The physical bases of the detection and characterisation of extrasolar 
planets  in the reflected light and thermal emission 
regimes are reviewed. They both have their advantages and disadvantages, 
including artefacts, in the determination of planet physical parameters (mass,
 size, albedo, surface and atmospheric conditions etc. A special attention
is paid for Earth-like planets and new perspectives for
these different aspects are also presented. 
\section{Introduction}
  The first discoveries of extrasolar planets 
have triggered a renewal of the permanent question
on the possible presence of life outside the Solar System. This question
can now be addressed in scientific  terms. Before detecting life on exoplanets,
it is necessary to detect and characterize these planets.
 I do here focus on the reflected light and thermal emission 
approaches of the direct imaging of planets. 
\section{Detection  of exoplanets by imaging}
Although  the most difficult, it is the most promising method for the 
characterization of planets. I will therefore
 remind its essential aspects. 
There are two kinds of emissions by a planet:
\begin{enumerate}
\item {\it Reflected light:}\\
       The planet reflects the stellar light with a flux ratio given by
\begin{displaymath}
     \frac{F_{refl}(t)}{F_*} =\frac{A_{pl}}{4}
     \times \left(\frac{R_{pl}}{a}\right)^2 \times \phi(t) ~~~ {\rm {(1)}}
\end{displaymath}
where $\phi(P,i,e,\omega,t)$ is an orbital phase factor (sinuso\"{\i}dal 
in case of
a circular orbit) and $A_{pl}$ the planet albedo. 
This ratio peaks at the same visible spectral range than 
the star itself and is typically $10^{-9}  - 10^{-10}$.
\item {\it Thermal emission:}\\
       The planet, heated by the star at a temperature
$T_{pl}=T_*\times (R_*/2a)^{1/2}(1-A_{pl})^{1/4}$, emits a thermal flux
given by
\begin{displaymath}
 \frac{F_{th}}{F_*}   = \left(\frac{R_{pl}}{2a}\right)^2  ~~~ {\rm {(2)}}
          \end{displaymath}
 This ratio peaks at the mid-infrared and  is typically $10^{-7}  - 10^{-7}$, 
{\it i. e.} about $10^3$ larger than in the visible range.
There is no orbital phase factor here. \\
Note that the above formulae {\it do not} hold for non spherical objects, 
{\it i.e.} planets with large Moons and planets with rings.
\end{enumerate}
\section{Application of imaging to the characterization of exoplanets}
Let us look with
more details into the planet characteristics accessible by imaging.

 {\it 1. Orbit:  }\\
 If an object is detected close to the star, one wants to determine its
orbit. Two orbital positions, together with the observation  epochs, 
{\it i.e.} 6 observables $t_1$, $t_2$, $x(t_1)$, $x(t_2)$, $y(t_1)$,
$x(t_2)$,  are in principle 
sufficient to determine the 6 orbital parameters $a$, $i$, $\omega$, $\Omega$,
$T_o$ and $e$. But one has to verify that the object is not a close background
star. This requires a third position measurement ($x(t_3)$, $y(t_3)$) at a 
third epoch  $t_3$.
  
This minimum
number at least holds for the thermal emission which is independent from
the orbital phase. For the reflected flux, one can take advantage of the
phase dependance given by (1) to reduce this number to 2 orbital positions
to deduce the orbit 
parameters. In that case it would indeed by unlikely that a 
background star had, by co\"{\i}ncidence, a flux variation precisely given 
by (1). This reduction from 3 to 2 of the number of  observations
of a star required
to assess the existence of a planet is important for missions 
scenarii and represents a significant advantage of  observations in the
 visible.

 {\it 2. Mass: } \\
In principle it can be determined only from the dynamical
        perturbation of the star's motion by the planet. Nevertheless, from
a low spectral resolution spectrum  (R=5) in the visible, one can
infer whether there is a high, medium or low density atmosphere.
From the latter, one can deduce, up to a factor 2-5
(Brown et al 2002), the mass of the planet
(low mass planets do not retain their atmosphere, while high mass planets
retain thick atmospheres). 
The thermal infrared is not suited for this type 
of studies. 

{\it 3. Radius:} \\
From formula (1) and from the fact that the albedo 
has an upper limit of 1, the visible flux gives an lower limit for the planet 
radius; unless a giant planet would have an albedo of 1\%, it cannot be
confused with an Earth-sized planet. The thermal emission gives, thanks to
the formula (2) a safer value for the radius (unless the planet is surrounded 
by Moons (DesMarais et al 2001) or by rings (Schneider 1999).

{\it 4. Temperature:} \\
For reflected light, it can be infered from
the star-planet distance and from the albedo through the relation
$T_{pl}=T_*\times (R_*/2a)^{1/2}(1-A_{pl})^{1/4}$. But then one has 
in principle to know the planet albedo. Nevertheless,  the latter
formula shows that the temperature is not very sensitive to the albedo:
a variation of $A_{pl}$ from 0.3 to 0.7 gives an decrease of 20\%
for $T_{pl}$. Here again, the thermal infrared gives a direct 
(and independent) measurement of $T_{pl}$, safer than from the reflected flux.

{\it 5. Albedo colour $A(\lambda)$:}   \\
The albedo can only be given by the
reflected flux. But, as seen on formula (1), only the product
$A_{pl}\times R_{pl}^2$ can be directly measured. Nevertheless, the 
measurement 
at different wavelengths gives the albedo {\it colour}, regardless of its 
absolute value. By itself this already constitutes a precious indication
on the nature of the planet surface, as shown by Brown et al. (2002).

 {\it 7. Environment: }
        \begin{itemize}
        \item {\it Atmosphere:}\\
                 As already mentionned, the albedo colour gives the amount of
   Rayleigh scattering, and thus the density of the atmosphere (Brown et al.
   2002).
        \item {\it Clouds:}\\
                The most natural explanation of
                chaotic variations of the albedo would be a variable cloud
                coverage. Let us note that a similar chaotic variation can 
                also be due to dust storms, like on Mars. In this case,
                 the confusion with clouds can be removed by the colour
                 characteristics of the albedo fluctuations: clouds have
                 a white albedo, while dust is red.
        \item  {\it Rings:}   \\
                Their existence would be inferred from a non Keplerian
                 variation of the phase factor $\phi(t)$. Indeed, its 
                 standard mathematical expression 
                ($\phi(t)= (1-\sin i \sin (2\pi t/P))/2$ in case of circular 
                orbits) holds only for spherical bodies. In presence of
rings, for half of the
                orbit, the observer sees only their backside, which is 
                black, giving to $\phi(t)$ a more complicated expression
                depending on the detailed configuration of the rings 
                 (Schneider 1999). This case is not an exception, as shown
               by the Solar System planets; it is quantitatively not 
              negligeable since for instance the reflected solar flux from
              Saturn rings is as large as the planet reflected flux itself.
        \item   {\it Moons   :}   \\
                 They will most likely be first detected by the transit method 
                (Sartoretti and Schneider 1999).
                 For the coming generation of imaging space missions
                ({\it e.g.} Darwin/TPF), the angular resolution will not
                 be sufficient to separate them angularly from their parent
                 planet. It will nevertheless be possible to detect them
                 by a photometric monitoring of the planet:\\
                  {\it a) Planet-satellite mutual transits} (Schneider   
                    2003).
                  A planet brightness drop with an amplitude
                  $(R_{sat}/R_{pl})^2$ should appear with a period
                   half the satellite revolution period. The geometric
                  probability of this event nevertheless does not exceed 
                   $\approx 10\%$. The event is detectable in both reflected
                   and thermal emission regimes.\\
                 {\it b) Planet-satellite mutual shadows} 
                  (Schneider   2003). 
                    It is most likely
                  that the satellite orbits lies close the planet orbital 
                  plane. In that case, the satellite throws, once per orbit,
                  a shadow on the illuminated part of the planet and, once 
                  per orbit, disappears in the planet shadow. An interesting
                   feature of this event is that
                  the satellite+planet flux drop has a very characteristic
                  shape. In case of a satellite orbit lying exactly in the 
                planet
                 orbital plane, this shape is,  for $\phi = \pi/2$,  
                  $\Delta F_{pl}/F_{pl}(\phi_{sat}) = \tan \phi_{sat}$, 
                  varying from 0 (when 
                  the satellite orbital  phase $\phi_{sat}=0$)
                   up to a maximum
                  $(R_{sat}/R_{pl})^{3/2}/\sqrt{2}$ which is larger by a 
                   factor $\sqrt{R_{pl}/(2R_{sat})}$ than the drop due to
                   mutual transits. 
                 For satellite orbits not lying exactly in the 
                planet orbital plane, the evolution of the function 
                   $\Delta F_{pl}(\phi_{sat})/F_{pl}$ along the planet 
                 orbital revolution 
                gives the two angular parameters characterizing the
                  relative inclination of the  planet and satellite orbital
                  planes. 
              In addition to being larger than mutual 
                      transits,
                  mutual shadows have a geometric probability close to 1.   
                  This event can be seen only for reflected light, {\it i.e.}
                  in the visible. 
         \end{itemize}
{\it 8. Surface properties:}
        \begin{itemize}
        \item {\it Structures: }\\
The formula (1) only gives  the product
$A_{pl}\times R_{pl}^2$. It thus does not enable to give the absolute value
of the planet albedo. But from the time variation of $F_{refl}(t)$ one can,
after correction of orbital effects, deduce the time variation $A_{pl}(t)$ of
the albedo (since the planet radius is constant). 
        A short term (hours to days) periodic variation
          would reveal the presence of surface inhomogeneities   of the albedo
     by the modulation of  $F_{refl}(t)$ due to the planet rotation. The period
       of the modulation gives the duration of the planet day, its
       amplitude gives the albedo contrast between different parts of the 
      planet surface (``continents'') and the shape of the modulation
      gives the spatial extension of ``continents'' (Schneider 1999, Ford, 
        Seager and Turner 2001).
         In principle the modulation of the
      thermal emission by oceans and continents could also be detected 
      during the diurnal
       planet rotation. But, while the continent/ocean contrast is about a
           factor 5 in reflected light, it is only 
       $4|T_{ocean}-T_{cont.}|/T_{mean}\approx 10\%$ for thermal emission.
        \item {\it Internal heat vs/ stellar heating:} \\
             Depending on the orbital phase, the observer
               sees the illuminated side or dark side of the planet.
          There may exist a temperature contrast $\Delta T_{pl}
           =  T_{pl,day}-T_{pl,night}$
            between these two sides. Along the orbital revolution
              it will provide an annual modulation of the effective
            planet temperature $T_{pl, eff}(t) \approx (T_{pl, day}^4(1-\sin i
            \cos (2\pi t/P))/2+T_{pl, night}^4(1-\sin i \sin (2\pi t/P))/2)
                ^{1/4}$ (for $e=0$). A low $\Delta T_{pl}$ would mean a
            high atmospheric or oceanic circulation, while a high
           $\Delta T_{pl}$ would mean a low lithsopheric heat conductivity.
       For instance, for the Earth the day/night
       temperature difference is about 10 K leading to a
            relative thermal flux variation $4(T_{pl,day}-
        T_{pl, night})/T_{mean}\approx 10\%$. For the Moon and Mars the 
              temperature
          difference is $\approx$ 100 K, giving a relative flux variation 
          of  a factor 2. The contrast $\Delta T_{pl}$ can be
             due to several factors: surface (lithosperic
                and oceanic) thermal conductivity, oceanic and 
             atmospheric circulation, and depends on the planet rotation rate.
             An additional source of thermal emission can be 
             purely internal,
             due to tectonic activity and to rocks radioactivity. It could
             in principle 
             produce a temperature in excess of the equilibrium
             temperature $T_{pl}=T_*\times (R_*/2a)^{1/2}(1-A_{pl})^{1/4}$.
             But the example of the Earth, for which the the tectonic and the 
            radiogenic heat flow is only 100 mW/m$^2$, compared to the
            $\approx$ 1 kW/m$^2$ heat flow produced by stellar heating,
               shows that this effect can be appreciable only for
               planets far away from their star, where the stellar heating
              is small. That is  {\it e.g.} the case of Io where the thermal 
                heating 
             causes about 12 one day volcanic outbursts per year 
             doubling the total 5 micron flux
                (Spencer and Schneider 1996).
             Together with the planet radius, mass (and thus the density), 
            age, albedo, the measurement of the effective temperature 
             and its 
            modulation will provide precious constraints on the planet
             atmospheric, surfacic and internal structure. Of course, this
              measurement is possible only in the infrared regime.
                      \end{itemize}
{\it 9. Life?  }\\
          A traditional prerequisit is the 
            presence of liquid water, imposing a planet temperature of about
          300 K. The planet must therefore lie in the ``habitable zone'',
          {\it i.e.} at a distance of
          $\approx (T_*/T_{\odot})^2(R_*/R_{\odot})$ AU from the star
          ($\approx$ from 0.1 to 1.5 AU, for M to F stars).\\
  The detection of signatures of Life (``biosignatures'') makes use
  of two approaches:
       \begin{itemize}
       \item {\it ``Dejecta'':} \\
               These are by-products of biological
        activity on the planet. The latter are mainly atmospheric gases such
         as O$_2$ (and its by-product O$_3$), CH$_4$. The key argument here
     is that on Earth all the molecular oxygen content of the atmosphere
        (20 \%)
        comes from the photosynthetic activity of vegetation and bacteria.
          This argument is enforced by the fact that on Mars and Venus
         there is no oxygen or ozone.
          Since the main
         source of carbon for organics is the atmospheric CO$_2$,
         the latter must also be present in the planet atmosphere.
                        The detection of O$_2$ is a priority in the sense
        that it gives an access to the  degree 
                   of biological  evolution on the planet (DesMarais et al.
         2001, 2002).                 O$_2$ is detectable only in the
        visible, all the other gases are detectable in both visible and
        infrared regimes (DesMarais et al.   2002).       
       \item  {\it ``Vegetation'':}   \\
              Whatever the detailed photosynthetic 
          mechanisms are, they must substract energy from some part
             of the stellar spectrum reflected by the planet, leading to 
         absorption features in this spectrum. This mechanism is responsible 
           for the ``red edge''  at 750 nm in the terrestrial vegetation
          spectrum. The latter has been observed globally, for the first time,
          for the whole Earth seen as an unresolved source in the Earthsine 
          spectrum (Arnold et al. 2002). The shape of this spectral feature
          gives some indication on the energy conversion mechanism, but
          the possible confusion with mineral absorption features has to be
           investigated further. It cannot be a safe biosignature by
                itself, it
          is useful only in association with other ones.
                \end{itemize}
The Table 1 summarizes the best wavelength regime for different planet
characteristics.

\begin{center}

Table 1\\

\bigskip 

\begin{tabular}{l|c|c}
{\large \bf Parameter} & {\large \bf Visible}&{\large \bf Infrared}\\ \hline
Radius & & yes\\
Mass & yes & \\
Temperature & yes & yes\\
Albedo & yes & \\
Day & yes& \\
Seasons  & yes& yes\\
Clouds& yes& \\
 Rings& yes& \\
Moons & &yes \\
O$_2$ & yes& \\
O$_3$, CH$_4$, CO$_2$, H$_2$O &yes & yes\\
Vegetation &yes & \\
Intern. heat& & yes\\
\end{tabular}
\end{center}

It seems that more science can be done with reflected light observations,
but it cannot do all of it and thermal infrared regime will
provide important complements.

\end{document}